
\input harvmac
\def\authornote{\footnote}
\hfuzz=15pt

\def\np#1#2#3{Nucl. Phys. {\bf #1} (#2) #3}
\def\pl#1#2#3{Phys. Lett. {\bf #1} (#2) #3}
\def\prl#1#2#3{Phys. Rev. Lett. {\bf #1} (#2) #3}
\def\pr#1#2#3{Phys. Rev. {\bf #1} (#2) #3}

\def\ie{{\it i.e.}}
\def\eg{{\it e.g.}}

\def\ccdot{\hbox{\kern-.1em$\cdot$\kern-.1em}}

\def\gu{\gamma^\mu}

 \def\sl#1{#1\hskip-0.5em /}  

\def\Asl{\hbox{/\kern-.6500em \rm A}}
\def\Dsl{\hbox{/\kern-.6000em\rm D}} 
\def\dsl{\,\raise.15ex\hbox{/}\mkern-13.5mu D} 

\Title{
  \vbox{
  \hbox{Brown HET-928}
  \hbox{hep-ph/9312353}
}}
{\vbox{
\centerline{Exact Heavy to Light Meson Form Factors}
\vskip1ex
\centerline{in the Combined}
\vskip1ex
\centerline{Heavy Quark, Large $N_c$ and Chiral Limits.}
}}

\centerline{Benjam\'\i n Grinstein\authornote{$^\star$}%
{Address until February 9, 1994.
Research supported in part by the Alfred P. Sloan Foundation and by the
U.S. Department of Energy under contract DE--AC35--89ER40486.
E-mail: {\tt ben@smuphy.physics.smu.edu}}
and Paul F.~Mende\authornote{$^{\star\star}$}
{Research supported in part by the TNRLC under an SSC Fellowship (\#FCFY9322)
and by the U.S. Department of Energy under grant DE-AC02-76-ER03130.
E-mail: {\tt mende@het.brown.edu}}}
\smallskip
\centerline{\sl $^\star$Superconducting Super Collider Laboratory,
Dallas, Texas 75237}
\smallskip
\centerline{\sl $^{\star\star}$Department of Physics, Brown University,
Providence, Rhode Island 02912}

\bigskip\bigskip

We demonstrate that the form factors of local operators between a heavy meson
state (like the~$B$) and a light pseudoscalar state (like the pion) are
given exactly by a single pole form in the combined heavy quark, large $N_c$
(number of colors) and chiral limits.
We discuss the deviations from this
exact result from finite heavy quark masses, non-zero light quark
masses and finite $N_c$.
We comment on
some of the numerous implications of this result.


\Date{December 1993}

\eject
\eject

\newsec{Introduction}
Little is known about form factors of local operators between
a heavy meson like the~$B$ --- with quantum numbers of
a single heavy quark~$Q$  and a
single light antiquark~$\bar q$ --- and light pseudoscalar mesons like the
$\pi$--$K$--$\eta$ octet.
Isgur and Wise have shown that heavy quark symmetries%
\ref\IWb{N.~Isgur and M.B.~Wise, \pl{B232}{1989}{113}\semi
N.~Isgur and M.B.~Wise, \pl{B237}{1990}{527}}
relate several form factors%
\ref\IWa{N.~Isgur and M.B.~Wise, \pr{D42}{1990}{2388}},
but nothing is known about their shape.
Thus far all theoretical attempts to describe them are based on
particular models of hadrons.

Surprisingly one can calculate the
shape of these form factors exactly in one
specific limit.
We show in this letter that when one takes
the leading term in a large number of colors
($N_c$) expansion and simultaneously takes the leading term in the heavy
quark expansion and the chiral limit, then the form factors are
a single pole {\it for all momentum transfers $t=q^2$},
\eqn\SHAPE{
   F(q^2) = { C\over q^2 - \mu^2 }
   + {\cal O}\left(1/N_c, \Lambda_{\rm QCD}/M_Q,
   \Lambda_{\rm QCD}/m_q\right)
.}
The location of the pole is $\mu^2 = \mu_{B^*}^2$, the squared-mass of the
heavy vector meson which couples to the heavy-light current,
\eqn\veccur{
   V^\mu=\bar u \gu b~.
}
The constant residue $C$ is completely determined in terms of the
decay constant $f_{B^*}$ and its coupling to a
$B$-$\pi$ pair, $g_{B^*\! B\pi}$.
To quantify the accuracy of this approximation
and the expansion around this
limit requires substantial exploration.
This is outlined below and further details
will be reported elsewhere
\ref\bgpfm{B.~Grinstein and P.F.~Mende,
in preparation}.

The most immediate application of this result is to the
decay
\eqn\DECAY{
   \bar B^0 \to \pi^+ \, e^- \, \bar\nu_e
,}
which is the direct route to obtaining the elusive
mixing angle $|V_{u b}|$ of the
CKM matrix of the standard electroweak theory
from measurements of $B$-meson decays.
Little data on this mode is available yet we may anticipate
the eventual measurement of the shape of its form factor
as the test of these ideas.

For a thumbnail preview of the discussion below, here is the
relevance of the three limits we consider:
(a) in the heavy-quark limit, including $1/M_Q$ corrections,
the $B$ and $B^*$ fall into a nearly degenerate $SU(2)$ multiplet;
(b) in the chiral limit, the conserved current mixes these states
and suppresses transitions between multiplets;
and (c) in the large-$N_c$ limit multiparticle intermediate states
are suppressed.

\nref\thooft{
G.~'t~Hooft,  \np{B72}{1974}{461}}

\newsec{Derivation of the form factor}

Consider the matrix element of a
local operator $\CO(x)$ with quantum numbers $Q$ and~$\bar q$:
\eqn\matrixelem{
   \vev{\pi(p')| \CO(0) | \bar B(p)}~.
}
For concreteness we shall speak specifically of
the $\bar B\to\pi$ transition.
In fact $\pi$ may generically stand for a light pseudoscalar meson,
conveniently thought of as a $q \bar q$ bound state; and $\bar B$ is a
ground state heavy meson, that is, the lightest with $Q\bar q$ quantum
numbers.
This matrix element can be written as a sum over tensor
structures times form factors, \ie, functions of the invariant
momentum transfer $q^2=(p-p')^2$. For example, the current
in~\veccur\ defines two form factors, $f_+$ and $f_-$:
\eqn\vecffs{
   \vev{\pi(p')| V_\mu | \bar B(p)}
   = f_+(q^2)(p+p')_\mu + f_-(q^2)(p-p')_\mu ~.
}
We evaluate the left-hand side by inserting a complete set of states
which couple with the same quantum numbers as the current.
In the large-$N_c$ limit \thooft\
the single-particle intermediate states dominate.
If $F(q^2)$ is a form factor, then in
the large-$N_c$ limit
\eqn\ffspoles{
F(q^2) = \sum_n {f_n g_{n B\pi}\over q^2 - \mu^2_n}
}
where the sum is over resonances~$B_n$ with masses~$\mu_n$,
and couplings $f_n$ and
$g_{n B\pi}$ to the current from the vacuum and to the $ B$--$\pi$ pair,
\eqn\gnbpi{
   g_{n B\pi} \propto \vev{\pi B_n | B}~.
}

Note that the sum over poles eq.~\ffspoles\
is very different from the statement of
eq.~\SHAPE\ that a single pole term contributes:
before specifying the residues the sum may
be a quite general function.
This generality is amply illustrated by the
case of $\bar Q Q$ charmonium-type states of heavy quarks
where the residues of many states are large and rapidly varying%
\ref\jaffemende{R.L.~Jaffe and P.F.~Mende, \np{B369}{1992}{189}}.
The massive poles are required by the structure of excited states,
yet confinement forbids the anomalous-threshold singularity
at small~$q^2$ which ought to be present to describe the
large size of a non-relativistic quark distribution.
Hence the electromagnetic form factor is given by exactly such a sum
of poles but with
{\it rapidly varying\/} numerator coefficients.
The form factor of heavy quarkonium is therefore
 never well-represented by a single term of~\ffspoles, yet is a smooth
function of~$q^2$ for all~$q^2< \mu_0^2$.

The lowest mass state in the sum~\ffspoles\
is either the ground state pseudoscalar meson~$B$ or the ground state
vector meson~$B^*$, according to whether the local operator $\CO$ has
odd or even parity, respectively.
It is obvious that this state dominates in the small kinematic region
\eqn\SMALLREGION{
   q^2 \approx \mu_B^2 ~;
}
to show that this state dominates {\it over a large range of~$q^2$\/}
is a dynamical question that must be addressed by evaluating
the behavior of the couplings~$f_n$ and~$g_{n B \pi}$.
This is the question which we take up here.

As usual, the states~$B$ and~$B^*$ have equal
masses in the leading order in the~$1/M_Q$ expansion;
the leading correction is the familiar hyperfine interaction
which introduces a spin-splitting,
$\mu_{B^*}-\mu_B=\CO(1/M_Q)$,
while
$\mu_{B^*}^2-\mu_B^2=\Lambda_0^2$
independent of~$M_Q$ in the large mass limit.

We wish to show that in the leading order in the~$1/M_Q$ expansion and
the chiral limit~$m_q\to0$ the couplings~$g_{n B\pi}$ vanish except for
the case when~$n$ corresponds to the~$B^*$ state.
Then the general expansion~\ffspoles\ reduces
to the form~\SHAPE.

Consider the how these same couplings arise in a different matrix element:
let us look at the {\it light quark\/} axial current,
\eqn\AXIAL{
   a_\nu = \bar q \gamma_\nu \gamma^5 q ,
}
between the~$B$ state and the generic resonance~$B_n$.
In the leading
order in the~$1/N_c$ expansion it is given as a sum
\eqn\vevofa{
   \vev{ B_n | a_\nu | B} = \sum_\ell
   {\vev{0|a_\nu|\pi^{(\ell)}}\vev{\pi^{(\ell)}B_n|B} \over
   p^{\prime2}-\mu_\ell^2}~.
}
where~$\ell$ runs over single particle states $\pi^{(\ell)}$
that are produced out of the vacuum by the light quark axial
current, and $\mu_\ell$ stands for its mass.

In the chiral limit only the pion, $\ell=0$, contributes to this sum
and the result is proportional to
the couplings of interest,
$\vev{\pi B_n|B}$.

To demonstrate this
let us show that, in the chiral limit,
\eqn\stepone{
   \vev{0|a_0|\pi^{(\ell)}(\vec p\,'=\vec 0)} = 0
}
except for the pion.
We need only consider the time component
of the current in the rest frame of the state.
In the chiral limit the axial current is conserved:
\eqn\proveone{
  \eqalign{ 0 &= \vev{0|\del_\mu a^\mu|\pi^{(\ell)}(\vec 0)} \cr
	&= p'_\mu \vev{0|a^\mu|\pi^{(\ell)}(\vec 0)} \cr
	&= \mu_\ell \vev{0|a_0|\pi^{(\ell)}(\vec 0)} \cr}
.}

Thus eq.~\vevofa\ reduces to
\eqn\vevofareducd{
 \vev{ B_n| a_0 | B} =
{i f_\pi p'_0 \vev{\pi B|n} \over p^{\prime2}}~,
}
when $\vec p\,'=0$.
Using conservation of $a_\mu$ again gives
\eqn\steptwo{
\eqalign{
	0 &= p'_0 \vev{ B_n| a^0 | B} \cr
	  &= i f_\pi \vev{\pi B_n|B} ~.\cr  }
}
This means that
\eqn\mevanishes{
   \vev{\pi B_n|B} = 0
}
in the frame $\vec p\,'=0$, but this holds generally since the matrix
element is a Lorentz scalar.

To go from $\vev{\pi B_n|B}$
to the form factors of $B\to\pi$ decay
we now must specify the spin-parity quantum
numbers of the state $|n\rangle$.
Consider first the case in which~$B_n$ is a scalar.
The off-shell matrix element $\vev{\pi B_n|B}$
can be characterized by a single `form-factor',
\eqn\scalarff{
   \vev{\pi B_n|B} = g^{(n)}(p'^2, q^2, p^2)~.
}
We apply a standard dispersion relation
to form factors~$f_\pm(q^2)$, which are of the form~\ffspoles,
to replace the matrix elements in the numerator by their
residue at the pole, $q^2=\mu_n^2$.
Thus to evaluate~$f_\pm$,  only the value of the form
factor on-shell is needed.
But we have just shown above that
\eqn\ZILCH{
   g^{(n)}(\mu_\pi^2,\mu_n^2,\mu_B^2) = 0 ~.
}
Hence scalar~$Q\bar q$ excited states do not contribute
to the resonant sum of eq.~\ffspoles.

Consider next what happens when~$B_n$ is a vector.
The matrix element is
\eqn\vectorffs{
\vev{\pi B_n|B} = (p+p')\cdot\epsilon\; g^{(n)}_+(q^2)
	+(p-p')\cdot\epsilon\; g_-^{(n)}(q^2)~,
}
where~$\epsilon$ is the polarization vector of the state~$B_n$
and its vacuum coupling through
the vector current is defined by
\eqn\vectovac{
   \vev{0|V_\nu|B_n} = i f_n \epsilon_\nu~.
}
The contribution to the $B\to\pi$ form
factors is then
\eqna\partialvecffs
$$
   \eqalignno{
   f_+ &= {f_n  g^{(n)}_-(\mu_n^2)\over q^2-\mu_n^2}
   &\partialvecffs a
   \cr
   f_- &= -f_n  g^{(n)}_+(\mu_n^2)/\mu_n^2
   - {f_n  g^{(n)}_-(\mu_n^2)\mu_B^2/\mu_n^2
   \over q^2-\mu_n^2}~. &\partialvecffs b
\cr}
$$
The vanishing of $\vev{\pi B_n|B}$, eq.~\mevanishes,
for an on-shell $B_n$ can now be
applied to the form factors $g_\pm$ and therefore to $f_\pm$.
Nothing is learned about~$g_+$ since $B_n$ is on-shell.
It is easy to see in the~$B_n$ rest frame
$\vec p = \vec p\,'$ that
\eqn\restgminus{
   g^{(n)}_- \vec\epsilon\cdot\vec p = 0
}
Let us restrict attention for now to the exact chiral limit
in the heavy quark (infinite) mass limit.
Then
$$
   \vec\epsilon\cdot\vec p  = |\vec p| \cos\theta ,
$$
where~$\theta$ is the angle between the polarization and the momentum
vectors, and is generally non-vanishing.
But from the kinematics it is also true that
\eqn\magofp{
 |\vec p| ={ \mu_n^2 - \mu_B^2 \over 2 \mu_n}
          =\cases{ \Lambda_0^2 / 2\mu_{B^*} = \CO(1/M_Q) & for $n = B^*$ ,\cr
                   \Lambda_n + {\cal O}(1/M_Q) & otherwise, \cr
		    }
}
where we introduce the mass difference $\Lambda_n\equiv \mu_n-\mu_B$ for
$n\neq B^*$ states, and take the large mass limit in the last equality.
Therefore, in the combined limit,
\eqn\FINAL{
   0 = g_-^{(n)}\, |\vec p| =
	 \cases{ 0 & for $n = B^*$ , \cr
		     g_-^{(n)}\Lambda_n & otherwise. \cr
		    }
}
The couplings to excited states thus go as
$g^{(n)}_- \sim {\cal O}(1/M_Q) \to 0$ except for $n=B^*$,
for which
$g^{(B^*)}_- \sim {\cal O}(1)$.

We thus obtain the advertised result that the form factor
is given by the single pole:
\eqn\FPLUS{
   f_+ = {f_{B^*} g^{(B^*)}_-(\mu_{B^*}^2)\over q^2-\mu_{B^*}^2}
   \approx f_- .
}
The last relation follows since $f_-$ satisfies
\eqn\FMINUS{
   f_- = -f_+ + \left[(1-\mu_B^2/\mu_{B^*}^2)f_+ + K \right],
}
where~$K$ is an undetermined constant.
Since we have taken the large
mass limit, our form factors should satisfy the standard scaling laws
\eqna\ffsscaling
$$
\eqalignno{
	f_+ + f_- &\sim \mu_B^{-1/2} &\ffsscaling a\cr
	f_+ - f_- &\sim \mu_B^{1/2} &\ffsscaling b\cr
	}
$$
and therefore the term in the square bracket in \FMINUS\ can
be neglected.

Some remarks:
\item{(1)}
The choices made of particular reference frames were for convenience
and not necessary to the derivation.
One can obtain, for instance, the same result~eq.~\mevanishes\
by taking the divergence on both sides of~\vevofa\
and letting $p'^2\to0$:
every term in the sum then vanishes except for the massless state.
\item{(2)}
Higher spin states can be readily incorporated into the discussion.  A
spin-$\ell$ meson is characterized by a totally symmetric traceless
transverse $\ell$-index ``polarization'' tensor,
$\epsilon_{\mu_1\ldots\mu_\ell}$.  The generalization of
eq.~\vectorffs\ has then, on the right hand, a sum over form factors
$$\sum_{\vbox{\hbox{\sevenrm sign permutations}}}
g_{\pm\cdots\pm}
\epsilon^{\mu_1\cdots\mu_\ell}
(p\pm p')_{\mu_1}\cdots (p\pm p')_{\mu_\ell}$$
As in the vector case, we only learn about one form factor, namely
$g_{-\cdots-}$. The rest of the argument is then just as before.
\item{(3)}
The large-$N_c$ limit suppressed multiparticle intermediate states
but not zero-particle intermediate states.
Any contact term contribution to the form factor shows up as
part of the constant~$K$ and is instead suppressed by a factor of
$1/M_Q$.

\newsec{Discussion}

We computed the hadronic form factor in the triple
limit of large-$N_c$, heavy quarks, and chiral symmetry.
We predict the weak decay form factor governing
$
   \bar B^0 \to \pi^+ \, e^- \, \bar\nu_e
$
to be pole-dominated as in the result eq.~$\FPLUS$
for all~$q^2$.
Moreover, the two factors in the numerator,
$f_{B^*}$ and~$g^{(B^*)}$, can be roughly estimated at least
by scaling measured values for charmed mesons using
heavy quark symmetry.

It is crucial to trace how the three limits ($1/N_c\to0$,
$1/M_Q\to0$ and $m_q\to0$) entered our derivation,
for the next goal must be to estimate the corrections
and understand the approach to the limit.
Is the result independent of the order of the limits?
If not, what limit and approximations are appropriate?

There is a subtle issue in the ordering of the approach to the
simultaneous $1/M_Q \to0$ and $m_q\to0$ limits.
In
ref.~\ref\IW{N.~Isgur and M.B.~Wise, \pr{D41}{1990}{151}},
for instance, Isgur and Wise examined the effect of the $B^*$ pole
on the $B\to \pi$ transition.
They concluded that pole-dominance held only
for a narrow kinematic region in contrast with our result.
They relied crucially on taking the pion mass to zero first
rather than, say,  holding $\mu_{\pi} \mu_B$ fixed.
Consider small chiral symmetry breaking corrections to
eq.~\mevanishes: for infinitesimal~$m_q$,
\eqn\mevanishesalmost{
   \vev{\pi B|B^*} \sim m_q \ph(M_Q,m_q,\Lambda) ,
}
where $\ph$ is some function of the masses and the hadronic
scale~$\Lambda$, and~$m_q\ph\to0$ as~$m_q\to0$.
For an on-shell $B^*$ this implies
\eqn\scalingalmost{
   g_-^{(B^*)}{\Lambda_0^2/M_Q} \sim m_q \ph(M_Q,m_q,\Lambda) ~.
}
Therefore, the function $\ph$ has a finite limit as $m_q \to0$ for
$M_Q$ fixed ($\Lambda$ is always fixed),
while $\lim_{M_Q\to\infty}\ph\sim 1/m_q$.

\nref\randall{
L.~Randall and M.B.~Wise, \pl{B303}{1993}{135}, hep-ph/9212315}
\nref\falkbg{A.~Falk and B.~Grinstein,
$\bar B\to\bar K e^+e^-$ in Chiral Perturbation Theory,
hep-ph/9306310, SLAC-PUB-6237, SSCL-Preprint-484, June 1993\semi
B.~Grinstein, \prl{71}{1993}{3067}
}

Examples of such functions are easy to come by.
A class of such functions is, for example,
\eqn\phexample{
   \ph(M_Q,m_q,\Lambda) \propto{ M_Q^{n-1}\over \Lambda^{n+1}+m_q M_Q^n}
}
for any~$n>1$.
The expansion parameter around the point~$1/M_Q=0$
is~$\Lambda^{n+1}/m_q M_Q^n$.
This type of expansion is familiar from calculations
in chiral perturbation theory for heavy mesons,
where one often finds\refs{\randall,\falkbg}\
corrections to be functions of
$
   \Lambda_0^2/\mu_\pi \mu_B\sim\Lambda_{\rm QCD}^{3/2}/m_q^{1/2}M_Q.
$
Just as the example eq.~\phexample,
the corrections computed in refs.~\refs{\randall,\falkbg}\
are not
singular in either of the limits $m_q\to0$ or $1/M_Q\to0$, but the
expansion parameters for the expansions about $m_q=0$ and $1/M_Q=0$
are the inverse of each other.
We see that the large mass and chiral
limits are inextricably coupled.

Let us consider one of these calculations in some detail.
Chiral
perturbation theory can be used to predict the leading corrections to
the form factors for semileptonic $B\to D$ or $D^*$ decays which are
generated at low momentum, below the chiral symmetry breaking
scale.
Deviations from the predicted normalization of form factors
arise from terms of order $1/M_Q^2$ in either the lagrangian or
the current as dictated by non-perturbative physics,
and there are
computable corrections that arise from the terms of order $1/M_Q$ in
the lagrangian which enter at one-loop.
Retaining only the dependence on the larger
hyperfine splittings $\Delta_D=\mu_{D^*}-\mu_D$,
the correction to the matrix elements at zero recoil
are\randall
\eqn\fromrandall{
\vev{D(v) | J^{\bar c b}_\mu|B(v)} = 2v_\mu\left(1-
{3g^2\over2}\left({\Delta_D\over4\pi f}\right)^2\left[ F(\Delta_D/\mu_\pi)
+\ln(\mu^2/\mu_\pi^2)\right] + C(\mu)/m_c^2\right)
}
where $C$ and stands for tree level counter-terms and
\eqn\fdefd{
   F(x)\equiv2\int_0^\infty d z {z^4\over (z^2+1)^{3/2}}
   \left({1\over[(z^2+1)^{1/2}+x]^2}- {1\over z^2+1}\right)
}
This matrix element has alternate, inequivalent expansions
around the limits
$x=0$ and $x=\infty$, which correspond to individually
taking $M_Q\to \infty$
and $\mu_\pi \to 0$, respectively.
For small~$x$, $F(x) \sim x$ while as $x\to\infty$,
$F(x)\sim \log x$ and this cancels the log singularity above.

It is instructive to compare the behavior of $g_-^{(B^*)}$ to
that of the form factors of higher vector states, $g_-^{(n)}$.
For these we expect a relation similar to \scalingalmost,
\eqn\scalingalmostmore{
   g_-^{(n)}{\Lambda_n} \sim m_q \ph(M_Q,m_q,\Lambda)~,
}
where $\ph$ is not necessarily the same function as above, but has the
same properties.
We therefore predict that at finite small~$m_q$
one should find
\eqn\SMALLM{
   g_-^{(n)}/g_-^{(B^*)}\sim 1/M_Q ~.
}

\nref\lukefalk{A. F. Falk and
M. Luke,  \pl{B292}{1992}{119}, hep-ph/9206241}

Consider the strong decays of the excited states $D_1'$ and $D_2^*$
into the $D$ or $D^*$ mesons and a pion studied in ref.~\lukefalk, as
well as the corresponding states where charm is replaced by bottom.
In the notation of ref.~\lukefalk\ these strong decays are described
in the combined heavy quark and chiral limits by the effective
(chiral) lagrangian
\eqn\afml{
\CL_{d}={h_1\over{\Lambda_\chi}}\Tr\left[\bar H_a T^\mu_b\left(i D_\mu\,
    \Asl\right)_{b a}{\gamma_5}\right]+{h_2\over{\Lambda_\chi}}
 \Tr\left[\bar H_a T^\mu_b\left(i\dsl A_\mu\right)_{b a}{\gamma_5}\right]
 + {\rm h.c.}\,,}
where $T^\mu$ and $H$ are the spin supermultiplets containing the
$D_1'$ and $D_2^*$, and $D$ and $D^*$ fields, respectively, and
$A_\mu$ is the axial vector field.  It follows that in the large-$N_c$
limit one must have~$h_1 + h_2 \sim \CO(1/M_Q) $.  This is nontrivial
information which is not automatically included in the effective
lagrangian formulation.

In contrast to the expectation that $g_{B^*\! B\pi}$ scales like
$M_Q$, it has been shown above that $g_{B^*\! B\pi}\sim 1$. Note that
this is necessary if the form factors in eq.~\FPLUS\ are to scale
according to \ffsscaling\null\ over the whole kinematic range. Recall
that $f_{B^*}\sim M_Q^{1/2}$, and consider the region of maximum
momentum transfer $q^2\approx q^2_{\rm max}\equiv(\mu_B-\mu_\pi)^2$.
In the chiral limit,
\eqn\fplusatmax{
   f_+(q^2_{\rm max})={f_{B^*}g_{B^*\! B\pi}\over\Lambda_0^2} + \cdots ,
}
where the dots indicate contributions of states above the~$B^*$.
As mentioned above, Isgur and Wise have argued that $f_\pm$ should be
pole-dominated in the proximity of $q^2_{\rm max}$ in the chiral limit\IW.
They observe that, were $g_{B^*\! B\pi}$ to scale like $M_Q$, the
$B^*$-pole contribution would scale like $M_Q^{3/2}$, while other
resonant contributions (and the continuum) would scale like
$M_Q^{1/2}$ even if their couplings to the $B$-$\pi$ pair were to
scale also as $M_Q$.
At least in the large-$N_c$ limit, this is not
the case at all.
$f_\pm$ never violate the scaling laws \ffsscaling\null,
and pole dominance occurs because the couplings to higher resonances
are suppressed by a power of~$M_Q$.

\nref\chirlag{G. Burdman, J. F. Donoghue, \pl{B280}{1992}{287}\semi
M. B. Wise,  \pr{D45}{1992}{2188}\semi
T.-M. Yan, H.-Y. Cheng, C.-Y. Cheung, G.-L. Lin,
Y. C. Lin and H.-L. Yu, \pr{D46}{1992}{1148}}

The large-$N_c$
expansion was used twice to write matrix elements as discrete sums
over single resonances, in eqs.~\ffspoles\ and~\vevofa.
The essential point was not so much the precise form of the sum
but rather
the absence, or suppression, of smooth background
contributions.
We do know that the large-$N_c$ limit
appears at least as good for heavy meson as for light mesons
\nref\BIGNBIGM{
B.~Grinstein and P.F.~Mende, \prl{69}{1992}{1018}, hep-ph/9204206;
M.~Burkardt and  E.S.~Swanson, \pr{D46}{1992}{5083}
}
\BIGNBIGM.

It is tempting to conjecture that $g_{B^*\! B\pi}\sim 1$ even at finite
$N_c$. This would avoid the contradiction with the scaling laws for
$f_\pm$. This scaling behavior of $g_{B^*\! B\pi}$ is important to
applications where the naive scaling is used to compare $B$ and $D$
meson couplings.
In the language of the effective chiral lagrangian of
ref.~\chirlag, the coupling $g$ of the $B$-$B^*$ multiplet to the pion
axial current scales like $1/M_Q$ rather than $M_Q^0$.

Both as a check on the assumptions discussed here and as a concrete
laboratory for exploring the expansion around the limit we have
analyzed the predictions of this work in the 't~Hooft model, QCD in
1+1 dimensions to leading order in the large--$N_c$ expansion.
Details of this work will be extensively discussed elsewhere\bgpfm.
In this exactly solvable model, we have shown that the single pole
picture is stable as one tunes the heavy quark mass down from infinity
and independently stable as one tunes the light quark mass up from
zero.
In two dimensions there is no spin, of course, so there are no
vector mesons.  Instead the $B$-meson couples directly to the vector
current and therefore plays the role of the $B^*$-meson of the
preceding discussion.  We show precisely that the couplings $g_{n
B\pi} $ vanish as $m_q\to0$ for $n\ne B$ but not for $n=B$.
Moreover, this single pole dominance holds for
any heavy quark mass~$M_Q$.

It is apparent that the result can be generalized in several
directions. For example, one may consider the form factor for a matrix
element of a local operator between a pion and non-ground state $Q\bar
q$ meson. Take, for example, the $B_1$ and $B_2^*$ mesons, which form
a multiplet of heavy quark spin symmetry. The above result states
that, modulo accidentally degenerate states, the form factors are pole
dominated by a pole at the $B_1$ or $B_2^*$ squared-mass. One may
consider other states for which the conserved axial current is a good
interpolating field, \eg, the $A_1$ pseudo-vector meson.  Perhaps more
interestingly, one may instead consider other conserved currents as
interpolating fields. The first obvious candidate is the vector
current $\bar q \gu q$. This can be used as interpolating field for
the vector mesons, like the $\rho$ and $K^*$.

{\bf Acknowledgments:}
We are indebted to Sidney Coleman for an incisive question and
comment which led to this work and we thank William Bardeen for
helpful discussions.

\listrefs
\bye